\documentclass[pra,nopacs,twocolumn,nofootinbib]{revtex4-1}
\usepackage{graphicx}
\setkeys{Gin}{draft=false} 
\usepackage{bm,amssymb,amsmath}
\usepackage{dcolumn}
\usepackage{natbib}
\usepackage
{hyperref}

\newcommand{\ecm}{\ensuremath{e {\cdotp} {\rm cm}}}
\newcommand{\de}{d_\mathrm{e}}
\newcommand{\eEDM}{{\em e}EDM}

\newcommand{\tref}[1]{Table~\ref{#1}}

\begin{document}
\title{Evaluation of CP-violation in HfF$^+$}

\author{A.N.\ Petrov$^{1,2}$}\email{alexsandernp@gmail.com}
\author{L.V.\ Skripnikov$^{1,2}$}
\author{A.V.\ Titov$^{1,2}$}
\homepage{http://www.qchem.pnpi.spb.ru}

   \affiliation{$^{1}$B.P.Konstantinov Petersburg Nuclear Physics Institute, Gatchina, Leningrad district 188300, Russia}
\affiliation{$^{2}$Dept.\ of Physics, Saint Petersburg State University, Saint Petersburg, Petrodvoretz 198504, Russia}

\author{V. V. Flambaum$^{3}$}
\affiliation{$^3$School of Physics, The University of New South Wales, Sydney
NSW 2052, Australia}

\date{\today, 20t}

\begin{abstract}
CP violation effects produced by  the nuclear magnetic quadrupole moment (MQM), electron electric dipole moment (EDM) and scalar$-$pseudoscalar nucleus$-$electron  neutral current (SP) interaction
in $^{177}$Hf$^{19}$F$^+$ and $^{179}$Hf$^{19}$F$^+$ are calculated. The role of the hyperfine interaction is investigated. 
It is shown that the MQM shift can be distinguished from the electron EDM and SP ones due to the implicit dependence of MQM shift on the hyperfine sublevel.
The MQM effect is expressed in terms of the proton (EDM), QCD vacuum angle $\theta$ and quark chromo-EDMs.
\end{abstract}

\maketitle


Recently, Cornell/Ye group has obtained the limit on the electron electric dipole moment (\eEDM), $|\de|<1.3\times 10^{-28}$ \ecm\ (90\% confidence), using trapped $^{180}$Hf$^{19}$F$^+$ ions \cite{Cornell:17} with the spinless $^{180}$Hf isotope. The measurements were performed on the ground rotational, $J{=}1$, level in the metastable electronic $H^3\Delta_1$ state. The experiment demonstrated a great potential for the investigation of the 
time-reversal and parity violating (T,P-odd)
 effects using HfF$^+$ ions \cite{Cornell:17, Petrov:18}. In Ref.~\cite{FDK14}  it was proposed to use $^{177}$Hf$^{19}$F$^+$ and $^{179}$Hf$^{19}$F$^+$ ions with isotopes of hafnium which have nuclear spin $I > 1/2$
 to measure T,P-odd  effects produced by the nuclear magnetic quadrupole moment (MQM). 
Measurement of the nuclear MQM is promising for establishing  new limits on neutron and proton  EDM, vacuum angle $\theta$ in quantum chromodynamics and EDMs and chromo-EDMs of the quarks \cite{Skripnikov:14a}.
Corresponding experiment is  discussed by Cornell/Ye group \cite{Ye2018}.

We have previously carried out theoretical studies of 
the effects induced by the nuclear MQM in $^{229}$ThO~\cite{Skripnikov:14a}, $^{181}$TaN~\cite{Skripnikov:15c} and $^{229}$ThF$^+$~\cite{Skripnikov:15b} molecules.
 In particular, 
we derived analytical expressions for the energy shifts of the molecule caused by the interaction of electrons with MQM, which can be measured in the experiment. However, the formulae do not take into account the mixing of different rotational levels by the 
magnetic dipole and electric quadrupole hyperfine interactions 
  and external electric field, which change the values of the shifts. 

The main goal of the present work is to perform numerical calculations of the MQM, \eEDM,  scalar$-$pseudoscalar nucleus$-$electron neutral current (SP) interaction energy shifts of different hyperfine sublevels in the ground rotational level of $^{177}$Hf$^{19}$F$^+$ and $^{179}$Hf$^{19}$F$^+$
which take into account these effects.

\section{T,P-odd interactions Hamiltonian}
\label{MHam}
Hafnium isotopes $^{177}$Hf and $^{179}$Hf have nuclear spin $I^1=7/2$ and $I^1=9/2$, respectively. Fluorine isotope $^{19}$F has nuclear spin $I^2=1/2$. For the purposes of this work it is convenient to use the coupling scheme
\begin{equation}
{\bf F}_1 = {\bf J} + {\bf I}^1,
\end{equation}
\begin{equation}
{\bf F} = {\bf F}_1 + {\bf I}^2,
\end{equation}
where {\bf J} is the total molecular 
less nuclear spins angular momentum.
The field-free energy levels of the ground rotational state with the quantum
number $J=1$ splits by the hyperfine interaction with the hafnium nucleus into
three groups which are characterized by $F_1=9/2$, $F_1=7/2$, $F_1=5/2$ for $^{177}$Hf and $F_1=11/2$, $F_1=9/2$, $F_1=7/2$ for $^{179}$Hf quantum numbers. The hyperfine interaction with the fluorine nucleus further splits levels with total momentum $F=F_1\pm1/2$.
Note that $F_1$ is not exact but a good quantum number since the
hyperfine interaction
with the fluorine nucleus is much weaker than the hyperfine interaction with the hafnium nucleus.
Finally each hyperfine level has two parity eigenstates known as the $\Omega$-doublet.
These states are equal mixture of the $\Omega=\pm1$ states, where $\Omega$ is the projection of ${\bf J}$ on the internuclear $\hat{n}$ axis.

An external electric field mixes $\Omega$-doublet states of opposite parity 
 and transforms the $\Omega$-doublet structure of each hyperfine level to the Stark doublet structure.
For a sufficiently large electric field, $\Omega$ becomes a good quantum number,
except $m_F=0$ levels which are not mixed by the electric field.
Here $m_F$ is the projection of ${\bf J}$ on the laboratory  $\hat{z}$ axis which coincides with the direction for the electric field.
The state $J=1$ in the  $^{180}$HfF$^+$ molecule with the spinless hafnium isotope becomes almost fully polarized by the  electric field $E > 4$ V/cm \cite{Cossel:12}. 
The molecules $^{177}$Hf$^{19}$F$^+$ and $^{179}$Hf$^{19}$F$^+$ require much larger field to be polarized (see below).

The T,P-odd electromagnetic interaction of the nuclear magnetic quadrupole moment with electrons
is described by the Hamiltonian \cite{Kozlov:87}
 \begin{align}
\label{hamq}
 H_{\rm MQM}  &=
 -\frac{  M}{2I^1(2I^1-1)}  T_{ik}\frac{3}{2r^5}\epsilon_{jli}\alpha_jr_lr_k,
 \end{align}
where $\epsilon_{jli}$ is the unit antisymmetric tensor, $\bm{\alpha}$ is the vector of Dirac matrices, $\bm{r}$ is the displacement of the electron from the Hf nucleus,  $M$ is the Hf nuclear MQM,
\begin{align}\label{eqaux1}
M_{i,k}=\frac{3M}{2I(2I-1)}T_{i,k}\, \\
 T_{i,k}=I^1_i I^1_k + I^1_k I^1_i -\tfrac23 \delta_{i,k} I^1(I^1+1) .
 \end{align}

The \eEDM\ interaction is described by the Hamiltonian
\begin{eqnarray}
  H_d=2d_e
  \left(\begin{array}{cc}
  0 & 0 \\
  0 & \bm{\sigma E} \\
  \end{array}\right)\ ,
 \label{Hd}
\end{eqnarray}
$\bm{E}$ is the inner molecular electric field, and $\bm{\sigma}$ are the Pauli matrices. 
 
The T,P-odd SP interaction with a characteristic dimensionless constant $k_{SP}$ is 
described by the Hamiltonian \cite{Hunter:91}
\begin{eqnarray}
  H_{SP}=i\frac{G_F}{\sqrt{2}}Zk_{SP}\gamma_0\gamma_5n(\textbf{r}),
 \label{Hsp}
\end{eqnarray}
where $G_F$ is the Fermi-coupling constant, $\gamma_0$ and $\gamma_5$ are the Dirac 
matrices 
and $n(\textbf{r})$ is the nuclear density normalized to unity. 
For simplicity the summation over different electrons is omitted in 
Eqs. (\ref{hamq}),(\ref{Hd}) and (\ref{Hsp}).

For the field-free eigenstates the expectation value for the T,P-odd operators 
(\ref{hamq}),(\ref{Hd}) and (\ref{Hsp}) is zero.
For the completely polarized molecule and neglecting the interaction between different rotational, $J$, and hyperfine, $F_1$,  levels the energy shift due to the SP, \eEDM and MQM interactions are \cite{Skripnikov:14a}
\begin{equation}
\label{shiftd}
\delta_d = d_eE_{\rm eff}\Omega,
\end{equation}
\begin{equation}
\label{shiftSP}
\delta_{\rm SP} = k_{\rm SP}W_{\rm SP}\Omega,
\end{equation}
\begin{eqnarray}
\label{shiftM}
\delta_M(J,F_1) = (-1)^{I^1+F}C(J,F_1) M W_M \Omega,\,\,\,\,\,\,\,\\
C(J,F_1)= \frac{(2J+1)}{2}
\frac{
    \left(
    \begin{array}{ccc}
    J &  2 &  J \\
   -\Omega & 0 & \Omega
    \end{array}
    \right)
    }
    {
    \left(
    \begin{array}{ccc}
    I^1 &  2 &  I^1 \\
   -I^1 & 0 & I^1
    \end{array}
    \right)
    }
    \left\{
    \begin{array}{ccc}
    J &  I^1 &  F_1 \\
    I^1 &  J & 2
    \end{array}
    \right\},\,\,\,\,\,\,\,\,
\end{eqnarray}
where $(...)$ means the elements with the 3j$-$symbols and $\{...\}$ with the 6j$-$symbols \cite{LL77}, 
\begin{equation}
\label{matrelem}
E_{\rm eff} =
\langle \Psi_{^{3}\Delta_1}|\sum_i\frac{H_d(i)}{d_e}|\Psi_{^{3}\Delta_1}\rangle,
\end{equation}
\begin{equation}
\label{WSP}
W_{\rm SP} =
\langle \Psi_{^{3}\Delta_1}|\sum_i\frac{H_{SP}(i)}{k_{SP}}|\Psi_{^{3}\Delta_1} \rangle,
\end{equation}
\begin{align}
\label{WM}
W_M= \frac{3}{2}\frac{1}{\Omega} 
   \langle
   \Psi_{^{3}\Delta_1}\vert\sum_i\left(\frac{\bm{\alpha}_i\times
\bm{r}_i}{r_i^5}\right)
 _\zeta r_\zeta \vert\Psi_{^{3}\Delta_1}
   \rangle,
 \end{align}
where $\Psi$ is the electronic wave function of the considered
$H^3\Delta_1$ state of the HfF$^+$ cation.

Two sublevels within a Stark doublet
are connected by the time reversal $m_F \rightarrow -m_F$, $\Omega \rightarrow -\Omega$
and therefore have opposite signs for both $m_F$ and $\Omega$ quantum numbers and are degenerate unless
the T,P-odd interactions are not taken into account.
One can see from  Eqs. (\ref{shiftd})-(\ref{shiftM})
 that the states with the opposite projections of $\Omega$ have the opposite MQM, \eEDM,  SP interaction energy shifts which give rise to a splitting between the Stark doublet sublevels to be measured in the experiment.
The MQM shift can be distinguished from 
the other two ones by its dependence on $F_1$ quantum number.
For given $F_1$ the energy shifts are independent of the 
$F$ and $m_F$ quantum numbers provided the approximations assumed above.
The exception is for $m_F=0$ levels which have zero shifts as they are equal mixture of $\Omega=\pm1$ states.

In the present work we take into account the interactions with the external electric field, 
magnetic dipole and electric quadrupole 
hyperfine interactions which mix different rotational ($J$) and hyperfine ($F_1$) levels and modify the values of the shifts given by 
Eqs.~(\ref{shiftd})-(\ref{shiftM}).

\section{Evaluation of T,P-odd shifts}
The MQM, \eEDM,  SP interaction energy shifts have been calculated as the expectation values of the 
Hamiltonians~(\ref{hamq}),(\ref{Hd}) and (\ref{Hsp}) 
with  the wavefunctions of the $^{177}$Hf$^{19}$F$^+$ and $^{179}$Hf$^{19}$F$^+$ molecules.
Following Refs. \cite{Petrov:11,Petrov:14}, the energy levels and wave functions of the  $^{180}$Hf$^{19}$F$^+$ ion are obtained by the numerical diagonalization of the molecular Hamiltonian (${\rm \bf \hat{H}}_{\rm mol}$)
on the basis set of the electronic-rotational wavefunctions
\begin{equation}
 \Psi_{\Omega}\theta^{J}_{M,\Omega}(\alpha,\beta)U^{\rm Hf}_{I^1M^1}U^{\rm F}_{I^2M^2}.
\label{basis}
\end{equation}
Here $\Psi_{\Omega}$ is the electronic wavefunction, $\theta^{J}_{M,\Omega}(\alpha,\beta)=\sqrt{(2J+1)/{4\pi}}D^{J}_{M,\Omega}(\alpha,\beta,\gamma=0)$ is the rotational wavefunction, $\alpha,\beta,\gamma$ are the Euler angles, $U^{\rm Hf}_{I^1M^1}$ and $U^{\rm F}_{I^2M^2}$ are the Hf anf F nuclear spin wavefunctions and $M$ $(\Omega)$ is the projection of the molecule angular momentum, {\bf J}, on the laboratory $\hat{z}$ (internuclear $\hat{n}$) axis, $M^{1,2}$ are the projections of the nuclear angular 
momentums on the same axis. 
We write the molecular Hamiltonian for the $^{180}$Hf$^{19}$F$^+$ molecule in the form:
\begin{equation}
{\rm \bf\hat{H}}_{\rm mol} = {\rm \bf \hat{H}}_{\rm el} + {\rm \bf \hat{H}}_{\rm rot} + {\rm \bf\hat{H}}_{\rm hfs} + {\rm \bf\hat{H}}_{\rm ext} .
\end{equation} 
Here ${\rm \bf \hat{H}}_{\rm el}$, ${\rm \bf\hat{H}}_{\rm rot}$, and ${\rm \bf\hat{H}}_{\rm ext}$ are the electronic, rotation of the molecule
and interaction of the molecule with the external field Hamiltonians, respectively, as they are described in Ref. \cite{Petrov:17b},
\begin{eqnarray}
 {\rm \bf\hat{H}}_{\rm hfs} = {\rm g}_{\rm F}{\mu_{N}} {\bf \rm I^2} \cdot \sum_i\left(\frac{\bm{\alpha}_i\times \bm{r}_{2i}}{{r_{2i}}^3}\right) + \\
{\rm g}_{\rm Hf}{\mu_{N}} {\bf \rm I^1} \cdot \sum_i\left(\frac{\bm{\alpha}_i\times \bm{r}_{1i}}{{r_{1i}}^3}\right) + \\
-e^2 \sum_q (-1)^q \hat{Q}^2_q({\bf \rm I^1}) \sum_i \sqrt{\frac{2\pi}{5}}\frac {Y_{2q}(\theta_{1i}, \phi_{1i})}{{r_{1i}}^3}
\end{eqnarray}
is the hyperfine interaction between electrons and a nucleus, ${\rm g}_F$ and ${\rm g}_{\rm Hf}$ are $^{19}$F and $^{177,179}$Hf nuclear g-factors,  $\mu_{N}$ is the nuclear magneton,
$\bm{r}_{1i}$ ($\bm{r}_{2i}$) is radius-vector for the $i$-th electron
 in the coordinate system centered on the Hf(F) nucleus,  $\hat{Q}^2_q({\bf \rm I^1})$ is the quadrupole moment operator
for the $^{177,179}$Hf nuclei.
Note, that the subscript $1$ is omitted in $r_1$ in the section \ref{MHam} for simplicity.
The hyperfine structure of the $^3\Delta_1$ state only was taken into account in the present study.
Provided that the {\it electronic} matrix elements are known, the matrix elements of ${\rm \bf\hat{H}}_{\rm hfs}$ between the states in the basis set (\ref{basis}) can be calculated with the help of the angular momentum algebra \cite{LL77}. The required {\it electronic} matrix elements are 
\begin{equation}
 \label{AllF}
   A^{\rm F}_{\parallel}={\rm g}_{\rm F}
   \langle
   \Psi_{^3\Delta_1}|\sum_i\left(\frac{\bm{\alpha}_i\times \bm{r}_{2i}}{r_{2i}^3}\right)_\zeta|\Psi_{^3\Delta_1} \rangle,
\end{equation}

\begin{equation}
 \label{AllHf}
   A^{\rm Hf}_{\parallel}={\rm g}_{\rm Hf}
   \langle
   \Psi_{^3\Delta_1}|\sum_i\left(\frac{\bm{\alpha}_i\times \bm{r}_{1i}}{r_{1i}^3}\right)_\zeta|\Psi_{^3\Delta_1} \rangle,
\end{equation}

\begin{equation}
 \label{Q0Hf}
   eQq_0=
   2eQ\langle
   \Psi_{^3\Delta_1}|\sum_i \sqrt{\frac{2\pi}{5}}\frac {Y_{20}(\theta_{1i}, \phi_{1i})}{{r_{1i}}^3}|\Psi_{^3\Delta_1} \rangle,
\end{equation}

\begin{equation}
 \label{Q2Hf}
   eQq_2=
   2\sqrt{6}eQ\langle
   \Psi_{^3\Delta_1}|\sum_i \sqrt{\frac{2\pi}{5}}\frac {Y_{22}(\theta_{1i}, \phi_{1i})}{{r_{1i}}^3}|\Psi_{^3\Delta_{-1}} \rangle,
\end{equation}
where $Q = 2\langle U^{\rm Hf}_{I^1I^1} |\hat{Q}^2_0({\bf \rm I^1})| U^{\rm Hf}_{I^1I^1}\rangle$ is the quadrupole moment for the $^{177,179}$Hf nuclei. The $^{177}$Hf  and $^{179}$Hf isotopes
have $I^1=7/2$, ${\rm g}_{\rm Hf}=0.2267$, $Q=3.365~{\rm b}$ and $I^1=9/2$, ${\rm g}_{\rm Hf}=-0.1424$, $Q=3.793~{\rm b}$, respectively. 
The magnetic dipole hyperfine structure constants $A^{\rm F}_{\parallel}= -62.0~{\rm MHz}$ was measured in Ref. \cite{Cornell:17}.
The magnetic dipole hyperfine structure constants $A^{\rm Hf}_{\parallel}=-1429~{\rm MHz}$ and $A^{\rm Hf}_{\parallel}=898~{\rm MHz}$ for $^{177}$Hf$^{19}$F$^+$ and $^{179}$Hf$^{19}$F$^+$, respectively, were calculated in Ref. \cite{Skripnikov:17c}.
The electric quadrupole hyperfine structure constants  $eQq_0= -2100~{\rm MHz}$, $eQq_2=  110~{\rm MHz}$  and 
$eQq_0= -2400~{\rm MHz}$, $eQq_2=  125~{\rm MHz}$ for $^{177}$Hf$^{19}$F$^+$ and $^{179}$Hf$^{19}$F$^+$, respectively, were calculated in the present work.
The ratios for the magnetic dipole and electric quadrupole hyperfine structure constants 
correspond to the ratios for the nuclear g-factors and the quadrupole moments of the
$^{177}$Hf and $^{179}$Hf nuclei.

\section{Evaluation of $eQq_0$ and $eQq_2$ constants}
To compute $eQq_0$ in the $^3\Delta_1$ state of HfF$^+$ we have performed relativistic coupled cluster calculations within the Dirac-Coulomb Hamiltonian using the {\sc dirac12} code \cite{DIRAC12}.
In all the calculations the Hf$-$F internuclear distance in the $^3\Delta_1$ state was set to 3.41 Bohr \cite{Cossel:12}.
In the main calculation all 80 electrons of HfF$^+$ were included in the correlation treatment within the coupled cluster calculations with single, double and perturbative triple amplitudes, CCSD(T) using the 
uncontracted
Dyall's CVTZ basis set for Hf \cite{Dyall:07,Dyall:12} and the ccpVTZ basis set \cite{Dunning:89,Kendall:92} for F. We have also applied the correction on the basis set expansion up to the 
uncontracted
Dyall's AEQZ basis set for Hf \cite{Dyall:07,Dyall:12} and the aug-ccpVQZ \cite{Dunning:89,Kendall:92} for F. In the calculation 
$1s-3d$ core electrons of Hf were excluded from the correlation treatment within the CCSD(T) method. Accounting of the perturbative triple cluster amplitudes contributes $\approx 88$ MHz in $eQq_0$($^{177}$HfF$^+$). The contribution of the higher cluster amplitudes were estimated within the two-step approach \cite{Petrov:02,Titov:06amin,Skripnikov:11a,Skripnikov:15b,Mosyagin:16,Skripnikov:16a}
similar to Refs.~\cite{Skripnikov:17b,Skripnikov:17a,Skripnikov:16b,Skripnikov:15c,Skripnikov:15a}
 and found to be negligible in the present case.

As it follows from Eq. (\ref{Q2Hf}),  $eQq_2$ has no non-zero matrix elements within a main nonrelativistic term $^3\Delta_1$.
The main contribution to $eQq_2$ is due to the spin-orbit admixture of $\Pi$ state with the leading configuration $|5s5d_1|$ composed
of $5s$, $5d$ atomic orbitals of Hf. Then one can obtain
\begin{equation}
 \label{Q2Hfa}
   eQq_2=483wQ\left<1/r^3\right>_{5d} {\rm MHz},
\end{equation}
where $w$ is the weight of the admixture of the $\Pi$ state, $Q$ is the quadrupole moment of $^{177,179}$Hf in Barns,  $\left<1/r^3\right>_{5d}$ =4.86 a.u.
as obtained from the Hartree-Fock-Dirac calculations of Hf$^+$.
The value for the $w$ can be estimated from the sensitive to it of the  body fixed g-factor
\begin{eqnarray}
 \label{Gpar}
       G_{\parallel} &=&\frac{1}{\Omega} \langle ^3\Delta_1 |\hat{L}^e_{\hat{n}} - {\rm g}_{S} \hat{S}^e_{\hat{n}} |^3\Delta_1 \rangle \approx \\
\nonumber
2 - 2.002319 + w,
\end{eqnarray}
where $ g_{S} = -2.002319$ is the  free$-$electron $g$-factor. The same approximation
as for Eq.(\ref{Q2Hfa}) was used in Eq.(\ref{Gpar}). Then using the experimental value $G_{\parallel}=0.011768$ 
\cite{Petrov:17b,Cornell:13}
one obtains $w=0.014$, $eQq_2=110$ MHz and $eQq_2=125$ MHz for $^{177}$Hf$^{19}$F$^+$ and $^{179}$Hf$^{19}$F$^+$, respectively.

\section{Results and discussions}

\begin{figure}
\includegraphics[width = 3.3 in]{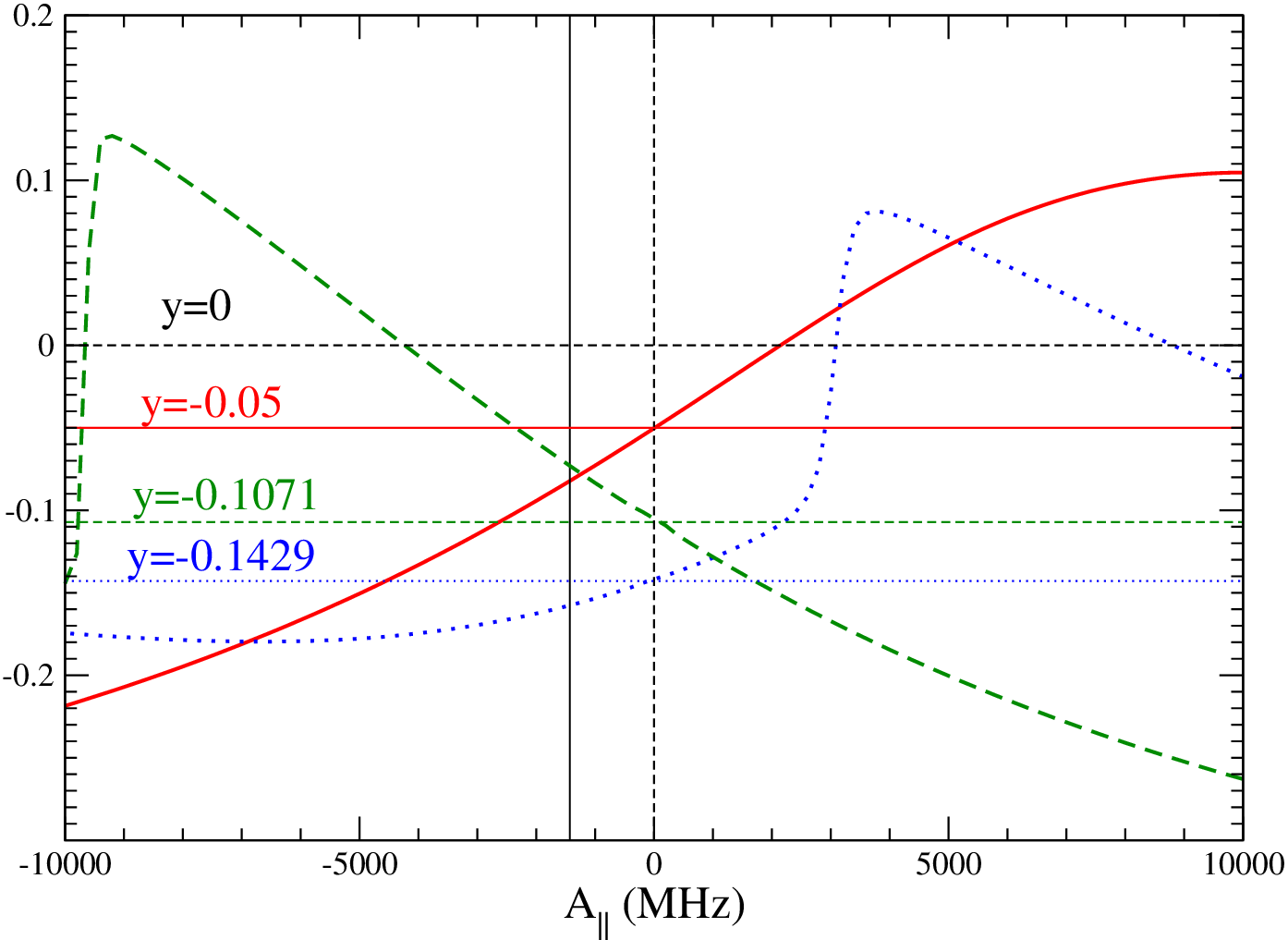}
\caption{(Color online) Calculated MQM energy shifts (in units $M W_M$) for the $J=1$, $H^3\Delta_1$ $^{177}$Hf$^{19}$F$^+$ as functions of $A^{\rm Hf}_{\parallel}$. Bold solid (red), bold dashed (green), bold dotted (blue) lines correspond to $F_1=9/2$, $F_1=7/2$, $F_1=5/2$, respectively. Horisontal thin solid (red), thin dashed (green), thin dotted (blue) 
correspond to values obtained by Eq. (\ref{shiftM}). Vertical black lines correspond
to values $A^{\rm Hf}_{\parallel}=0$ and $A^{\rm Hf}_{\parallel} = -1429$ MHz obtained in calculation. $E=20$ V/cm, $eQq_0 = 0$ and $eQq_2 = 0$ in calculations.
\\
\\
}
 \label{fApar}
\end{figure}

\begin{figure}
\includegraphics[width = 3.3 in]{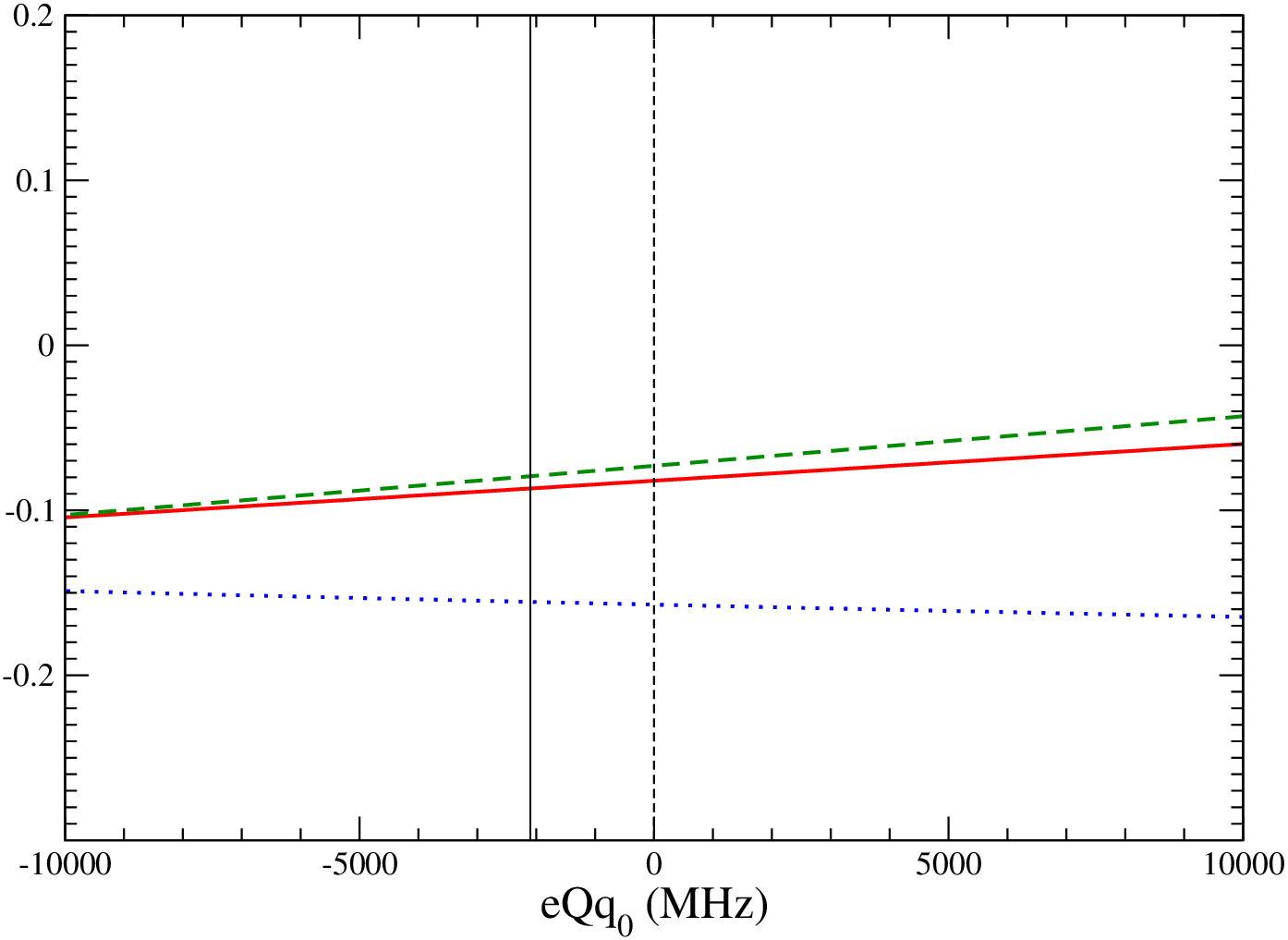}
\caption{(Color online) Calculated MQM energy shifts (in units $M W_M$) for the $J=1$, $H^3\Delta_1$ $^{177}$Hf$^{19}$F$^+$ as functions of $eQq_0$. Bold solid (red), bold dashed (green), bold dotted (blue) lines correspond to $F_1=9/2$, $F_1=7/2$, $F_1=5/2$, respectively. 
Vertical black lines correspond
to the values $eQq_0 = 0$ and $eQq_0 = -2100$ MHz obtained in the calculation. $E=20$ V/cm, $A^{\rm Hf}_{\parallel} = -1429$ MHz and $eQq_2 = 0$ in the calculations.
\\
\\
}
 \label{feQq0}
\end{figure}

\begin{figure}
\includegraphics[width = 3.3 in]{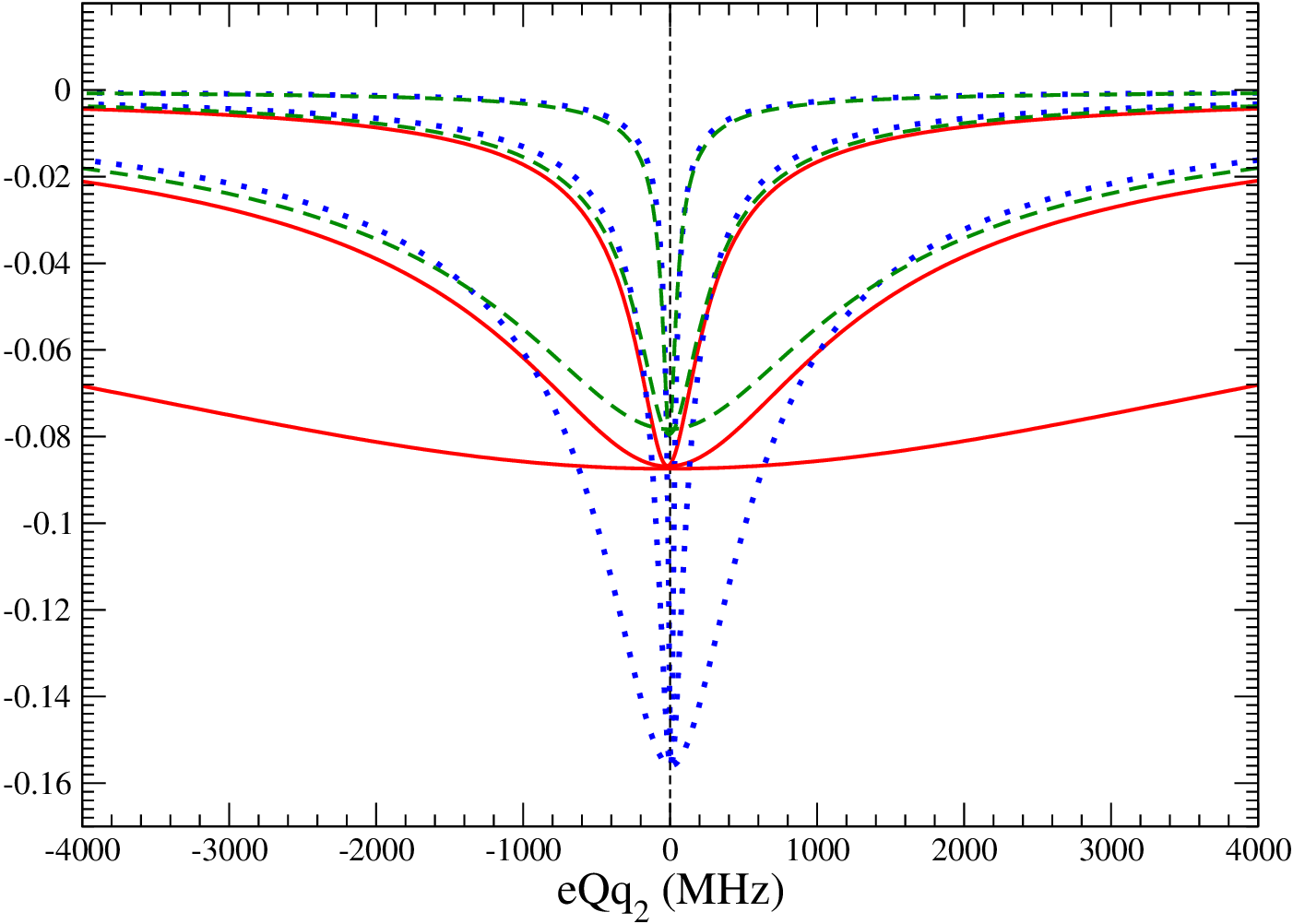}
\caption{(Color online) Calculated MQM energy shifts (in units $M W_M$) for the $J=1$, $H^3\Delta_1$ $^{177}$Hf$^{19}$F$^+$ as functions of $eQq_2$. Bold solid (red), bold dashed (green), bold dotted (blue) lines correspond to $F_1=9/2$, $F_1=7/2$, $F_1=5/2$, respectively. 
$E=4,20,100$ V/cm, $A^{\rm Hf}_{\parallel} = -1429$ MHz and $eQq_0 = -2100$ MHz in calculations.}
 \label{feQq2}
\end{figure}

In Figs. (\ref{fApar})-(\ref{feQq2})
 results for MQM shifts as functions of $A^{\rm Hf}_{\parallel}$, $eQq_0$ and $eQq_2$ are given for $^{177}$Hf$^{19}$F$^+$.
One 
can see
that MQM shift strongly depends on $eQq_2$ and decreases as $eQq_2$ increases. Similar dependencies are for the SP and \eEDM\ shifts. The reason is that the electric quadrupole hyperfine interaction causes the sublevels $\Omega = +1$ and $\Omega=-1$, 
which have different signs for T,P-odd shifts, to mix. The coupling of the states with the different signs of $\Omega$ is proportional to $eQq_2$ as it
follows from Eq. (\ref{Q2Hf}). The least dependence is observed for the $eQq_0$ constant. Hamiltonians
 (\ref{Hd}) and (\ref{Hsp})
  do not mix different
rotational levels and therefore are almost independent of the 
 $A^{\rm Hf}_{\parallel}$ and $eQq_0$ constants. 
In Table (\ref{TResultShift}) the calculated T,P-odd shifts accounting for the hyperfine and Stark interactions between different rotational levels compared with the ones obtained from 
Eqs. (\ref{shiftd})-(\ref{shiftM})
 are given. The values are quite different and the former demonstrate a large dependence on the electric field.

For the completely polarized HfF$^+$ the \eEDM\ and SP interaction energy shifts approach $d_eE_{\rm eff}^a$ and $k_{\rm SP}W_{\rm SP}$ values, respectively, which are independent of the 
F$_1$ quantum number. As evident from 
\tref{TResultShift}  the MQM shifts depend on F$_1$. This fact has to be used
to distinguish MQM from the scalar T,P-odd effects. Eq. (\ref{shiftM}) gives approximately a factor of three difference between the largest and the smallest shifts
for both $^{177}$Hf$^{19}$F$^+$ and $^{179}$Hf$^{19}$F$^+$ ions.
One can see
from \tref{TResultShift} that more accurate numerical calculations
give approximately factors two and four for $^{177}$Hf$^{19}$F$^+$ and $^{179}$Hf$^{19}$F$^+$, respectively. 
MQM can be expressed in terms of the 
proton and neutron EDMs d$_p$ and d$_n$, QCD vacuum angle $\theta$, quark chromo-EDMs ${\tilde d_u}$ and  ${\tilde d_d}$. Using these data \cite{FDK14}, the calculated value $W_M = 0.494 \frac{10^{33}\mathrm{Hz}}{e~{\rm cm}^2}$ \cite{Skripnikov:17b}, and data from \tref{TResultShift} one obtains that current limits
$|\rm{d}_p| < 2.0\cdot10^{-25}e~{\rm cm}$, $|\theta| < 1.5\cdot10^{-10}$, $|{\tilde d_u}-{\tilde d_d}| < 5.7\cdot10^{-27}e~{\rm cm}$ correspond to energy
shifts less than 6$\mu$Hz, 8$\mu$Hz, 18$\mu$Hz for $^{177}$Hf$^{19}$F$^+$ and
less than 3$\mu$Hz, 5$\mu$Hz, 11$\mu$Hz for $^{179}$Hf$^{19}$F$^+$,
respectively.

\section{Conclusion} 
We have calculated the T,P-odd energy shifts produced by the  MQM, \eEDM, and  SP interactions for the ground rotational $J = 1$ hyperfine levels of the $^3\Delta_1$ electronic state of the $^{177}$Hf$^{19}$F$^+$ and $^{179}$Hf$^{19}$F$^+$ ions. It is found that taking into account the hyperfine interaction is critically important for the accurate evaluation of the effects. The MQM shifts depend on a hyperfine sublevel of the 
$J = 1$ rotational state. We found that 
there is factor of two and four difference between the largest and the smallest shifts for $^{177}$Hf$^{19}$F$^+$ and $^{179}$Hf$^{19}$F$^+$, respectively.
The experiment on $^{177}$Hf$^{19}$F$^+$ and $^{179}$Hf$^{19}$F$^+$, similarly to the one on $^{180}$Hf$^{19}$F$^+$ \cite{Cornell:17}, can be performed using rotating electric and magnetic fields which trap the cation. We have shown that $^{177}$Hf$^{19}$F$^+$ and $^{179}$Hf$^{19}$F$^+$ require much larger than  $^{180}$Hf$^{19}$F$^+$ electric field to be polarized.
However, our calculations have shown that the F$_1 =9/2, 5/2$ and F$_1 =11/2, 7/2$ states of $^{177}$Hf$^{19}$F$^+$ and $^{179}$Hf$^{19}$F$^+$ respectively, for the electric filed $\sim 20$ V/cm used in the experiment \cite{Cornell:17}, are almost completely polarized.

\begin{table*}[!h]
\caption{
$\delta_M$, $\delta_d$ and $\delta_{\rm SP}$ shifts (in units $M W_M$, $d_eE_{\rm eff}^a$ and $k_{\rm SP}W_{\rm SP}$, respectively) for the $J=1$, $H^3\Delta_1$. 
Numerical calculations take into account interaction between different rotational levels.
}
\label{TResultShift}
\begin{tabular}{ c  d  d  d  d}
\hline
$F_1$ & {\rm Electric}       &    \multicolumn {2}{c}{$\delta_M$}                  &  \delta_d (\delta_{\rm SP})             \\
      &  \rm field (V/cm)    & \multicolumn {1}{c}{\rm Eq.  (\ref{shiftM})$^b$ }  & \rm numerical    &    \rm numerical          \\
      &                      &                                                    &  \rm calculation &  \rm   calculation^c      \\
\hline
\multicolumn {5}{c} {$^{177}$Hf$^{19}$F$^+$}  \\
\hline
 9/2  &        4             &      0.05000       &           0.07879    &        0.90804    \\
      &       20             &                    &           0.08651    &        0.99577    \\
      &      100             &                    &           0.08742    &        0.99983    \\
 7/2  &        4             &      0.14286       &           0.02278    &        0.14581    \\
      &       20             &                    &           0.09268    &        0.59325    \\
      &      100             &                    &           0.15050    &        0.96511    \\
 5/2  &        4             &      0.10714       &           0.02864    &        0.36015    \\
      &       20             &                    &           0.07048    &        0.88791    \\
      &      100             &                    &           0.07789    &        0.99466    \\
\hline
\multicolumn {5}{c} {$^{179}$Hf$^{19}$F$^+$}  \\
\hline
11/2  &        4             &      0.05000       &          0.02165    &        0.72080     \\
      &       20             &                    &          0.02961    &        0.98200     \\
      &      100             &                    &          0.03074    &        0.99926     \\
 9/2  &        4             &      0.13333       &          0.00446    &        0.03696     \\
      &       20             &                    &          0.02195    &        0.18187     \\
      &      100             &                    &          0.08213    &        0.67962     \\
 7/2  &        4             &      0.09167       &          0.06467    &        0.53716     \\
      &       20             &                    &          0.11499    &        0.95405     \\
      &      100             &                    &          0.12101    &        0.99803     \\
\hline
\end{tabular}
\\
$^a$  $\delta_d$ and  $\delta_{\rm SP}$ are equal.  \\
$^b$ Values obtained using Eq. (\ref{shiftM}) are independent of 
the values of the electric field. \\
$^c$ If the hyperfine interaction with the hafnium nucleus is not taken into account (or we consider $^{180}$Hf$^{19}$F$^+$ ion where it is identically zero) then the $\delta_d$ and  $\delta_{\rm SP}$  shifts are equal to unity for the fields used in the table.
\end{table*}

\section{Acknowledgement}
The formulation of the problem and the molecular calculations 
are supported by the Russian Science Foundation grant No. 18-12-00227.
The electronic structure calculations of $eQq_0$ parameter 
were supported by RFBR, according to the research Project No.~16-32-60013 mol\_a\_dk.
The electronic structure calculations were performed at 
the PIK data center of NRC ``Kurchatov Institute'' -- PNPI.
The calculations of the nuclear structure 
are supported by the Australian Research Council and New Zealand Institute for Advanced Study.


\bibliographystyle{./bib/apsrev}


\end{document}